# A General Method to Incorporate Spatial Information into Loss Functions for GAN-based Super-resolution Models

Xijun Wang, Santiago López-Tapia, Alice Lucas, Xinyi Wu, Rafael Molina, *Life Senior Member, IEEE* and Aggelos K. Katsaggelos, *Life Fellow, IEEE*

*Abstract*—Generative Adversarial Networks (GANs) have shown great performance on super-resolution problems since they can generate more visually realistic images and video frames. However, these models often introduce side effects into the outputs, such as unexpected artifacts and noises. To reduce these artifacts and enhance the perceptual quality of the results, in this paper, we propose a general method that can be effectively used in most GAN-based super-resolution (SR) models by introducing essential spatial information into the training process. We extract spatial information from the input data and incorporate it into the training loss, making the corresponding loss a spatially adaptive (SA) one. After that, we utilize it to guide the training process. We will show that the proposed approach is independent of the methods used to extract the spatial information and independent of the SR tasks and models. This method consistently guides the training process towards generating visually pleasing SR images and video frames, substantially mitigating artifacts and noise, ultimately leading to enhanced perceptual quality.

## I. INTRODUCTION

Super-resolution (SR) is a traditional image/video processing task aiming at estimating high-resolution (HR) images or videos from low-resolution (LR) ones. Researchers have introduced diverse methodologies and algorithms to address this task [1]. In recent years, Deep Learning (DL) has emerged in the field, exhibiting exceptional performance in addressing diverse image processing challenges, including SR [2], [3], [4], [5], [6]. In pioneering works using deep neural network (DNN) on SR problems, Dong et al. [7], [8] developed the SRCNN for single image super-resolution (SISR). This approach learned end-to-end mapping from a single LR image to its HR counterpart using a Convolutional Neural Network (CNN). Following this approximation, Kappeler et al. [9] designed the VSRnet, a three-layer CNN for video super-resolution (VSR). Both works have outperformed the previous state-of-art methods. After that, various DNN architectures have been created in these fields [10], [11]. More recently, the emphasis in recent studies within this field has shifted from merely achieving accurate image or video recovery to generating visually pleasing outcomes. To achieve this objective, generative adversarial networks [12] have been widely used in current state-of-art studies [13], [14], [15], [16], [17], [18], [19], [20], [21], [22]. These works demonstrated that the adversarial learning strategy has the capability to steer generative outcomes toward the realm of natural images, a realm perceptually authentic to humans. Consequently, this approach yields images and video frames with a realistic appearance.

The majority of outcomes produced by GAN architectures have surpassed those of non-GAN architectures in terms of perceptual quality for images and video frames. To further enhance perceptual quality, recent studies have endeavored to incorporate specific spatial information into the network's training process. This selected spatial information often plays a pivotal role in enhancing an image's visual quality, encompassing elements like edges and textures. For example, Jiang et al. [23] proposed an edge-enhancement subnetwork to recover the high-frequency edge details of images. Wang et al. [24] and Wu et al. [25] integrated a Spatial Feature Transform layer into their network's architecture. This layer leveraged semantic texture information conditioned on semantic segmentation maps during training. Zhao et al. [26] designed a region-level non-local module and integrated it into the generative network to capture long-range dependencies between features. These aforementioned studies collectively demonstrate that appropriately incorporating essential spatial information during GAN network training can significantly contribute to generating visually pleasing outcomes. However, these methods rely on modifying or constructing new architectures or layers within their networks. Such approaches often possess limitations due to their strong dependence on specific network architectures, hindering straightforward adaptation to different network structures. In such cases, using the same approach may not reach the same performance, and many adjustments will be needed. In the current works, few have tried to include the spatial information by improving the objective loss, which can be easily extended to different models regardless of their architectures, as similar loss functions are employed during

This work has been submitted for possible publication. Copyright may be transferred without notice, after which this version may no longer be accessible.

This work was supported by Sony Research Program.

X. Wang is with the Department of Computer Science, Northwestern University, Evanston, IL 60208 USA.

S. López-Tapia, Alice Lucas, X. Wu, and A. K. Katsaggelos are with the Department of Electrical and Computer Engineering, Northwestern University, Evanston, IL 60208 USA.

R. Molina is with the Computer Science and Artificial Intelligence Department, Universidad de Granada, 18071 Granada, Spain.

training. Therefore, this paper proposes an effective and efficient approach for enhancing the training loss through the integration of spatial information. This is achieved by adapting the original loss function to become spatially adaptive (SA). We will show we can generate better images and video frames with higher perceptual quality after training different networks with the SA losses. Because of the superior performance of GAN in the SR field, in this paper, we primarily focus on GAN models.

In state of art GAN-based super-resolution models [13], [14], [15], [16], [17], [19], [23], [24], [22], [21], [26], training losses always contain two primary components: adversarial loss and distance-based fidelity losses. The adversarial loss originates from the adversarial mechanism, initially introduced by Goodfellow et al. [12]. This loss drives the adversarial training dynamics between the generator and discriminator networks. The distance-based fidelity losses are commonly defined in pixel or feature spaces, and they can be included for different purposes. For example, most of them are used to assess the results' spatial quality [13], [14], [15], [16], [17], [19], [23], [24], [21], [26], which compute the difference between super-resolved images/frames and their corresponding HR counterparts in the low-level pixel space or the high-level feature space. While some of them are computed between adjacent estimated video frames, aiming at improving the results' temporal quality in the video super-resolution scenarios [17]. In this paper, since our goal is to further improve the results' perceptual quality, our focus lies on the losses used for discriminating the spatial quality. As previously discussed, high-frequency details such as edges are pivotal for human vision. However, these details are usually more challenging to be accurately recovered than the flat areas, and more distortions and artifacts tend to be exhibited around these edge areas. Such distortions and artifacts can be easily observed by human eyes and negatively impacts the results' perceptual quality as well as the accuracy of subsequent vision applications. Therefore, this paper selects edge information as the representative of spatial information. We first extract spatial (edge) information from input images or video frames. We then incorporate it into the original distance-based fidelity loss (defined in the pixel space in particular), making the new training loss a spatially adaptive(SA) one. We will show that the proposed SA loss can effectively guide the network to pay more attention to recovering those edge areas and generate a more accurate and sharper edge structure. This augmentation significantly enhances the visual impact of the ultimate super-resolved outcomes. Besides, instead of adding additional loss terms, the SA loss idea is proposed by using the spatial information in the commonly-used pixel loss to train the SR DNN models. It is also independent of the network's architecture, thus can be easily and efficiently incorporated along with different network architectures.

The rest of the paper is organized as follows. In Section II, we provide an overview of related works. Section III shows the methods employed for extracting the spatial (edge) information from input data. Then, we explain how to use the extracted spatial information to construct the SA distance-based pixel loss. Section IV shows the effectiveness of the proposed SA loss idea used in GAN-based models. To do so, we we choose two prominent and extensively employed SISR and VSR models as representatives, replace their original pixel loss with the SA pixel loss and retrain the models. After that, from both quantitative and qualitative aspects, we will show that models trained with the SA loss consistently generate sharper and clearer edges with fewer distortions and artifacts, therefore producing more visually satisfactory results than the same models trained with their original loss. In Section V, we draw our conclusion and outline potential directions for future research.

## II. Related Work

During the initial stages of DL method development to address SR tasks [7], [8], [9], researchers commonly adopted the distance-based pixel loss for network training. This loss quantifies the distance between the resultant super-resolved image and the corresponding high-resolution (HR) ground truth image, evaluated within the low-level pixel space. This pixel loss tends to be very effective, and many of the well-known models in the literature like VDSR[10], EDSR[11], RCAN[27], CARN[28], EDVR[29] and so on only used such pixel-wise loss to train their networks.

Nevertheless, since the pixel loss cannot effectively take the image's perceptual quality into account, the generated images/video frames are often far from perceptual satisfaction. In order to evaluate the perceptual quality of generated images, Johnson et al. [30] first introduced the perceptual loss (also called content loss or feature loss in some works) and used it in training the SISR network. This loss measures the semantic difference between high-level image feature representations extracted from pre-trained image classification networks. In contrast to pixel loss, perceptual loss encourages the output image to be perceptually consistent with the target image instead of forcing them to match pixels only. With perceptual loss included, the trained networks can produce visually pleasing images. Therefore, it is widely used in most state-of-art works pursuing the high perceptual quality [13], [14], [15], [16], [17], [19], [23], [24], [30], [31], [21], [26], [22].

Benefiting from the exceptional learning capability of GANs [6], post-adversarial training equips the resultant generator to yield outputs conforming to the distribution of authentic data. Ledig et al. [8] first proposed the SRGAN and got more photo-realistic outputs. Since then, an increasing number of works in this field have chosen to build their model based on GAN, thereby incorporating the adversarial loss as part of their training loss [13], [14], [15], [16], [17], [19], [23], [24], [31], [22].

The testing outcomes have demonstrated that even though the SR models trained with the perceptual loss and adversarial loss achieved lower PSNR results than those trained solely with pixel loss, they exhibit significant increases in results' perceptual quality [15], [16], [22]. Hence, GAN-based models will serve as the foundational models for this study.

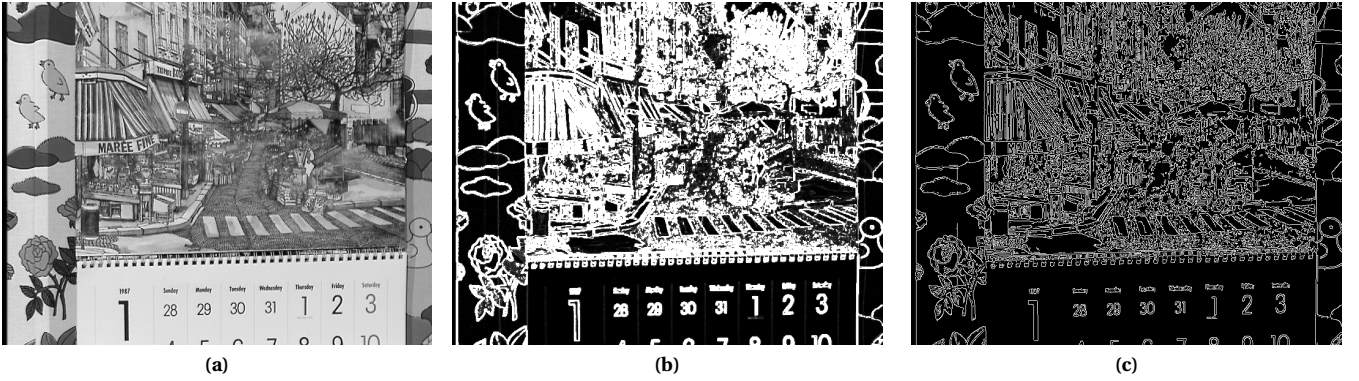

**Figure 1:** The gray image and its edge maps (a) The gray image; (b) Edge map computed by local variance; (c) Edge map computed by Canny detection.

Besides the commonly-used losses we have introduced above, some additional distance-based losses have also been used to train the SR models for different purposes. However, accumulating too many losses for the training is not always advantageous. In practice, people often form the final training loss function by aggregating a weighted summation of these losses to constrain the generation process in different aspects [13], [14], [15], [16], [17], [19], [23], [24], [31], [32], [33], [34], [21], [22]. For example, when training the GAN-based models, in order to balance the perception-distortion trade-off, people often need much empirical exploration to find suitable weights for different losses [14], [15], [16], [17], [19], [32], [33], [34], [21], [22]. Therefore, in this study, instead of adding additional loss terms, we present a methodology to directly enhance the fundamental and prevalent pixel loss by incorporating spatial information, thus transforming it into a spatially adaptive pixel loss. We illustrate that the proposed spatially adaptive (SA) loss contributes to a further enhancement of the output's visual quality. Additionally, since the proposed SA loss is defined in the pixel space, it can be easily used in the Non-GAN models as well.

## III. Proposed Method

Edge information is a crucial factor that influences image and video quality. Precise and accurate edges can improve the perceptual feeling of super-resolved images and video frames. Therefore, we select edge information as the primary representative of spatial characteristics. We will first extract edge information from the input data, then propose our methodology of using it into the low-level pixel training loss space, making the conventional pixel loss to be the SA one. Such SA loss can guide the models to pay more attention to recovering edge areas during training, preserving the edge information as much as possible in the end. Our experiments demonstrate that employing the proposed SA loss in the training of both image SR GAN and video SR GAN models results in a noticeable enhancement in the visual quality of the output images and video frames.

### A. Extracting Spatial information

As explained above, we use edge information as our spatial information. To extract edges from the input images or video frames, we utilize two different algorithms in this paper: one is the local variance algorithm, producing an edge mask with values ranging from 0 to 1, and the other one is the well-known canny edge detection algorithm, yielding a binary edge mask with values of 0 or 1. We use two different edge extraction algorithms here because we hypothesized that the proposed SA loss idea is independent of the edge extraction algorithms or the edge masks employed. As long as the algorithm can efficiently extract the edge information from the input data, the proposed SA loss can work successfully using the corresponding extracted edge mask. In the following, we will first describe the local variance and canny edge detection algorithms that we have used.

In this paper, we view the images as multi-channelized. For example, the gray image has one channel, the RGB-color image has three channels, and the images in the feature domain commonly have multiple channels.

1) Local variance edge detection: The local variance edge detection algorithm [35] takes a multi-channel image $x$ with element $x(k, i, j)$ as input, where $k$ represents the channel index and $(i, j)$ is the pixel location within the $k$th channel. The local variance $\mu_{k,i,j}(x)$ at pixel location $(i, j)$ within the kth channel is computed by:

$$\mu_{k,i,j}(x) = \sum_{(l_1, l_2) \in \Gamma_{k,i,j}} \frac{1}{|\Gamma_{k,i,j}|} (x(k, i+l_1, j+l_2) - m_{k,i,j}(x))^2, \quad (1)$$

where

$$m_{k,i,j}(x) = \sum_{(l_1, l_2) \in \Gamma_{k,i,j}} \frac{1}{|\Gamma_{k,i,j}|} x\left(k, i+l_1, j+l_2\right) \quad (2)$$

is the local mean. $\Gamma_{k,i,j}$ is the analysis window around $x(k, i, j)$, and $|\Gamma_{k,i,j}|$ denotes the number of elements in the analysis window.

The local mean $m_{k,i,j}(x)$ and local variance $\mu_{k,i,j}(x)$ compute the average and the variance of the pixels within the analysis window $\Gamma_{k,i,j}$. It is evident that the local variance $\mu_{k,i,j}(x)$ exhibits high values in high-frequency image areas such as edges, while displaying low values in flat regions. We further normalize its values to the range [0,1], getting the final edge map, defined as $W(x)$, with elements

$$w_{k,i,j}(x) = \frac{\mu_{k,i,j}(x)}{\mu_{k,i,j}(x) + \delta}, \tag{3}$$

where $\delta > 0$, a tuning parameter determined experimentally. It is easy to see that in flat regions $w_{k,i,j}(x) \approx 0$, while in areas of high spatial activity like edge regions $w_{k,i,j}(x) \approx 1$. An example of the edge map computed by the local variance method for a grayscale image ($k = 1$) is shown in Figure 1(b) (The displayed edge map image is a result of scaling $W(x)$ by 255). We can see that the edge areas take higher values than the non-edge(flat) areas.

2) Canny edge detection: Since canny edge detection [36] is a well-known and typical edge extraction method, we won't further explain its details here. When designing the experiments, we directly use the canny function implemented in the OpenCV library [37]. The canny method outputs a binary edge map, where the edge pixels have a value of 1, and the non-edge pixels have a value of 0. The corresponding edge map derived through the Canny method is illustrated in Figure 1(c). Again, the edge map image shown there is scaled by 255, and we can also see that the edge areas take higher values than the non-edge (flat) areas.

Next, we are going to show how we include the extracted edge information into the training losses of the SR models, making them the SA ones.

### B. Spatially Adaptive pixel loss

In the current SR DL models, almost all use distance-based losses as their training loss or part of their training loss. In particular, the SR GAN models nowadays tend to use distance-based losses to regularize the GAN training, and these losses are often defined in pixel and feature spaces [14], [15], [16], [17], [19], [24], [25], [34], [21], [26], [22]. Our goal is to use the spatial information (edge information in this work) extracted from the input data into the distance-based loss defined in the pixel space. We name this enhanced pixel loss the spatially adaptive (SA) pixel loss. It can guide the model to generate clearer and more accurate edges, ultimately leading to the creation of visually captivating images and video frames. In the following, we will use two prevalent distance-based loss formats – the L1 norm and the Charbonnier norm as examples to show how to fulfill this goal. The loss functions of the SISR and VSR GAN models used in this paper follow these two loss formats, which we will show in the next subsection. The idea can be easily used for other distance-based loss formats in the same way.

The L1 norm distance loss is defined as:

$$l^1(u, v) = \sum_k \sum_i \sum_j |(u_{k,i,j} - v_{k,i,j})|, \tag{4}$$

and the Charbonnier distance loss is defined as:

$$\gamma(u, v) = \sum_k \sum_i \sum_j \sqrt{(u_{k,i,j} - v_{k,i,j})^2 + \epsilon^2}, \tag{5}$$

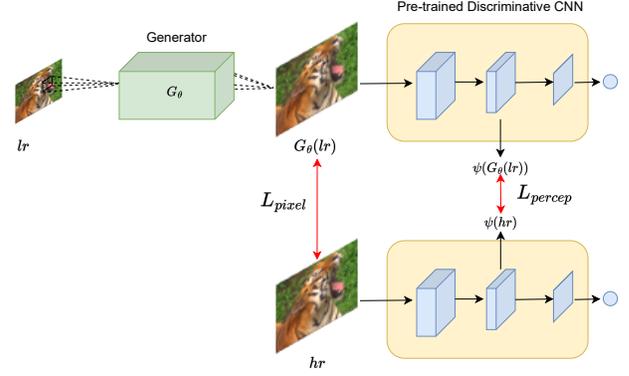

**Figure 2:** Visualization of pixel loss and perceptual loss.

for the general case of two multi-channel images u and v, with elements $u_{k,i,j}$ and $v_{k,i,j}$ respectively. Index $k$ is the channel number, e.g., $k = 1$ for a gray-scale image, $k \in \{1, 2, 3\}$ for a color image, and $k \in \{1, 2, 3...\}$ for an image defined in the feature space in the DL settings. $\epsilon$ is a small constant close to zero, for example, $\epsilon = 0.001$.

The idea for distance losses is straightforward, involving the computation of dissimilarity between two images. In the context of constructing the training loss for super-resolution deep learning (SR DL) models, these losses serve to quantify differences between the synthesized SR images and ground truth HR counterparts, and the computation is often conducted within both pixel and feature spaces. Consequently, the objective of the SR models centers on the minimization of these losses, thus driving the generation of SR images to closely resemble their corresponding HR counterparts.

One limitation of the definition of the above distance-based losses is that all elements in the images are weighted equally. However, we would like to preserve the edge information in the SR images as much as possible. Towards this end, during training the models, regions of high spatial activity areas (edge areas) should be weighted heavier than the smooth regions. Thus, we propose the following modification of the L1 and Charbonnier distance-based losses:

$$l_w^1(u, v, H) = \sum_k \sum_i \sum_j h_{k,i,j}|(u_{k,i,j} - v_{k,i,j})|, \tag{6}$$

$$\gamma_w(u, v, H) = \sum_k \sum_i \sum_j h_{k,i,j}\sqrt{(u_{k,i,j} - v_{k,i,j})^2 + \epsilon^2}, \tag{7}$$

where $H$ is a weight matrix, including the set of weights $h_{k,i,j}$, it weighs the difference between $u$ and $v$ at position $(k, i, j)$.

The following will show how to include the spatial activity information into the pixel loss by using the above-modified L1 and Charbonnier losses.

Using Equation 6 and 7, the original pixel losses for training the SR models are used to be defined as:

$$L_{pixel}^{l1} = l_w^1(hr, sr, \alpha 1), \tag{8}$$

$$L_{pixel}^{\gamma} = \gamma_w(hr, sr, \alpha 1), \tag{9}$$

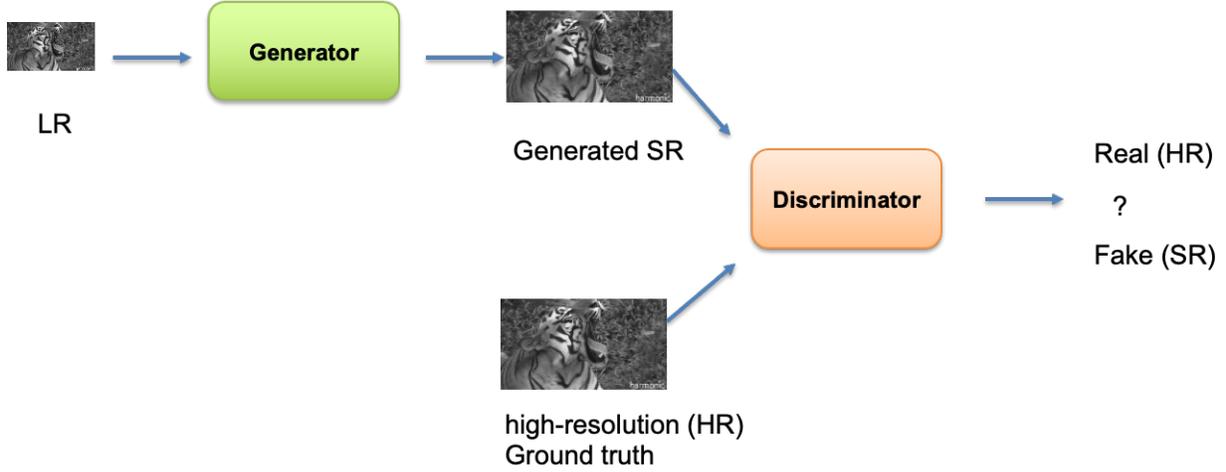

LR

Generated SR

Real (HR)

?

Fake (SR)

high-resolution (HR)
Ground truth

**Figure 3:** Super-resolution with GANs.

where $hr$ is the HR image, and $sr$ is the corresponded SR image. $\mathbf{1}$ is a weight matrix whose elements are all 1. The coefficient $\alpha$ is a hyper-parameter, determining the weight of the pixel loss term. Such hyper-parameters balance different loss terms in the final training loss. As previously mentioned, the original pixel loss functions uniformly treat all image elements, but we want to stress the edge areas during training. We then propose the spatially adaptive (SA) pixel losses which bring the edge information in, defined as:

$$L_{SA-pixel}^{l1} = l_w^1(hr, sr, \beta W(hr)), \tag{10}$$

$$L_{SA-pixel}^{\gamma} = \gamma_w(hr, sr, \beta W(hr)), \tag{11}$$

where $W(hr)$ is the weight matrix described in Section III-A, i.e., the edge map we extracted from the input HR image $hr$ using either local variance or canny methods, and its elements are weights $w_{k,i,j}$. Recall what we have learned above about $W(hr)$: in areas of high spatial activity like edge regions, we have $w_{k,i,j}$ equals 1 or close to 1, depending on which edge extraction methods we choose. In contrast, in the flat areas, we have $w_{k,i,j}$ equals 0 or close to 0. Therefore, the values of $w_{k,i,j}$ in the edge regions are larger than the values in the flat regions. The larger the value of $w_{k,i,j}$ is, the more important the corresponding pixel (at position $(k, i, j)$) becomes in the function to be optimized. Because the difference calculated from these edge pixels will contribute a larger loss value to the total loss than the difference computed from the smooth pixels. Therefore, to minimize the loss value, during training, our models will consider these edge pixels as more important pixels and put more effort into recovering them. In other words, during the backward pass, the trainable parameters will be updated "consciously" towards the direction where more accurate edges can be super-resolved, and larger weight updates will be given to those parameters responsible for super-resolving these edge-like regions. Again, $\beta$ is a hyper-parameter, serving as the coefficient of the SA pixel loss used in the final training loss.

However, we still have one problem left for the SA pixel losses defined in Equation 10 and 11. When we are using the edge map $W(hr)$ as our weight matrix, in the flat areas, we have $w_{k,i,j}$ equals 0 or close to 0, which means we almost completely ignore to super-resolved the flat regions, and this is not appropriate. Therefore, in practice, we consistently retain a certain level of original pixel loss when forming the SA pixel loss, so the final SA pixel losses we used are:

$$L_{SA-pixel}^{l1} = l_w^1(hr, sr, \alpha\mathbf{1} + \beta W(hr)), \tag{12}$$

$$L_{SA-pixel}^{\gamma} = \gamma_w(hr, sr, \alpha\mathbf{1} + \beta W(hr)), \tag{13}$$

and the reason for including the $\alpha\mathbf{1}$ term is to ensure that the loss does not ignore flat regions in the images.

### C. GAN Loss and Perceptual Loss

The GAN-based super-resolution model engages two networks: a generator and a discriminator. As shown in Figure 3, the generator takes LR images as input and generates the super-resolved image. The discriminator takes an image as input, analyzing input images to distinguish between images generated by the generator (i.e., fake images) and those from the ground truth high-resolution dataset (i.e., real images).

A minimax game ensues between the generator and discriminator, where the generator endeavors to generate super-resolved images capable of deceiving the discriminator, letting the discriminator perceive them to be real images. The discriminator strives to not be fooled by the generator and discriminates the fake images from the real ones.

Adapting the GAN formulation first introduced in Goodfellow et al.'s work [12] to super-resolution tasks involves addressing the adversarial min-max problem:

$$\min_\theta \max_\phi L_{GAN}(\phi, \theta) = \mathbb{E}_x \left[\log D_\phi(hr)\right] + \tag{14}$$
$$\mathbb{E}_Y \left[\log\left(1 - D_\phi(G_\theta(lr))\right)\right],$$

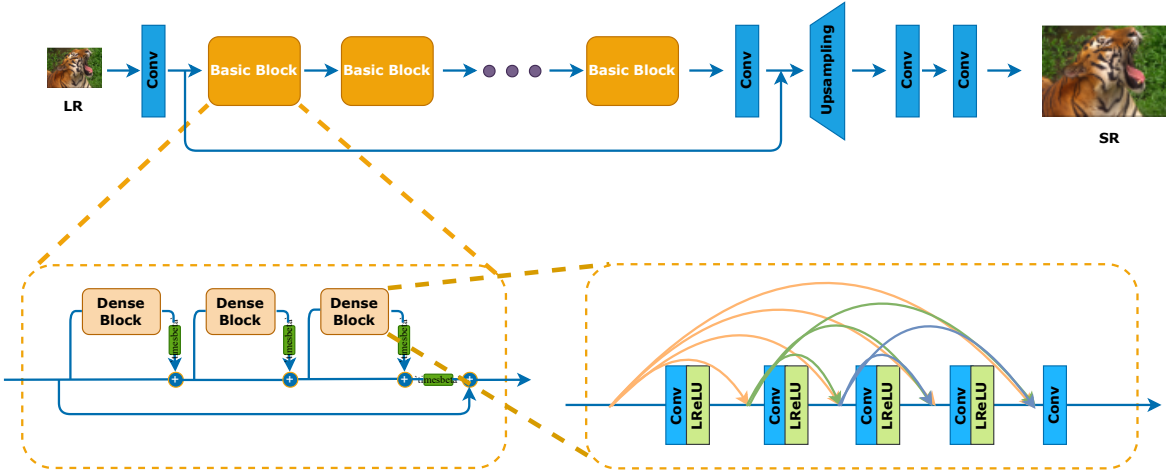

**Figure 4:** The ESRGAN's generator architecture. It consists of a convolutional (conv) operation applied to the input image, followed by 23 basic blocks and a conv layer. A long skip connection is used, where the output feature maps will add with the output feature maps from the first conv operation. After this, up-sampling and two more conv operations are conducted to obtain the final SR image output. The basic block used here is Residual-in-Residual-Dense-Block (RRDB), which combines the multi-level residual network and dense connections. [15]

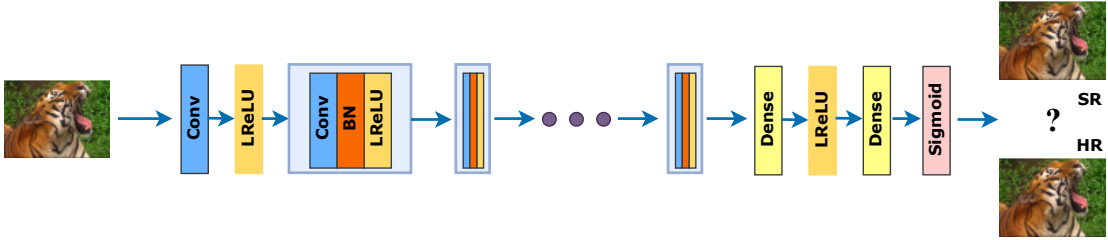

**Figure 5:** The ESRGAN's discriminator architecture. It consists of conv, Leaky ReLU, and batch normalization layers along the way, followed by two fully connected layers and a sigmoid activation function to obtain a probability.[15]

where $lr$ denotes the LR input, $D_\phi$ refers to the discriminator with trainable parameters $\phi$, and $G_\theta$ represents the generator network with trainable parameters $\theta$. Therefore, $G_\theta(lr)$ represents the output of the generator, which corresponds to the super-resolved image $sr$.

Because the GAN-based model could be unstable during training and is not easy to converge appropriately. Typically, training the generator can be more challenging than training the discriminator. Therefore, in addition to the adversarial loss, two other distance-based losses, namely the pixel and perceptual loss, are commonly employed when training the generator, as depicted in Figure 2.

We have defined the pixel loss, which calculates the distance between the generated SR image and the HR image in pixel space. Regarding the perceptual loss, it measures the distance between the generated SR image and the HR image in a perceptual space, which is defined by the middle layer output of a pre-trained discriminative CNN model when taking the HR and SR as inputs, respectively:

$$L_{percep} = \gamma(\psi(hr), \psi(G_\theta(lr))), \tag{15}$$

where the feature space denoted as $\psi(\cdot)$ is computed from the activations of the intermediate layers of the $VGG$ network [38]. Without loss of generality, we adopt the

Charbonnier distance format $\gamma$, which can be easily adapted to other distance formats.

### D. Spatially Adaptive loss used in Super-resolution models

The spatially adaptive loss can be effectively used to improve the results regardless of the chosen SR models. In this paper, we use two widely utilized SR GAN models as illustrative instances to apply the SA loss, one for image SR (ESRGAN [15]), the other for video SR (VSRResFeatGAN [16]). They both adopted the GANs framework and used perceptual loss during training. We will show that enhancing their initial training loss with the SA loss leads to a subsequent augmentation in the visual quality of the output results. The results comparison will be shown in Section V.

*1)* SA loss for single image super-resolution model: ESRGAN adopts the GAN framework. Accordingly, the model comprises two networks: a generator and a discriminator. The generator's architecture is shown in Figure 4. It takes an LR image as input and outputs the corresponding SR image. The discriminator's architecture is shown in Figure 5. Its input is either an HR image or an SR image generated from the generator, and its output is the probability of whether the input is a genuine HR image or otherwise. More details can be found in [15].

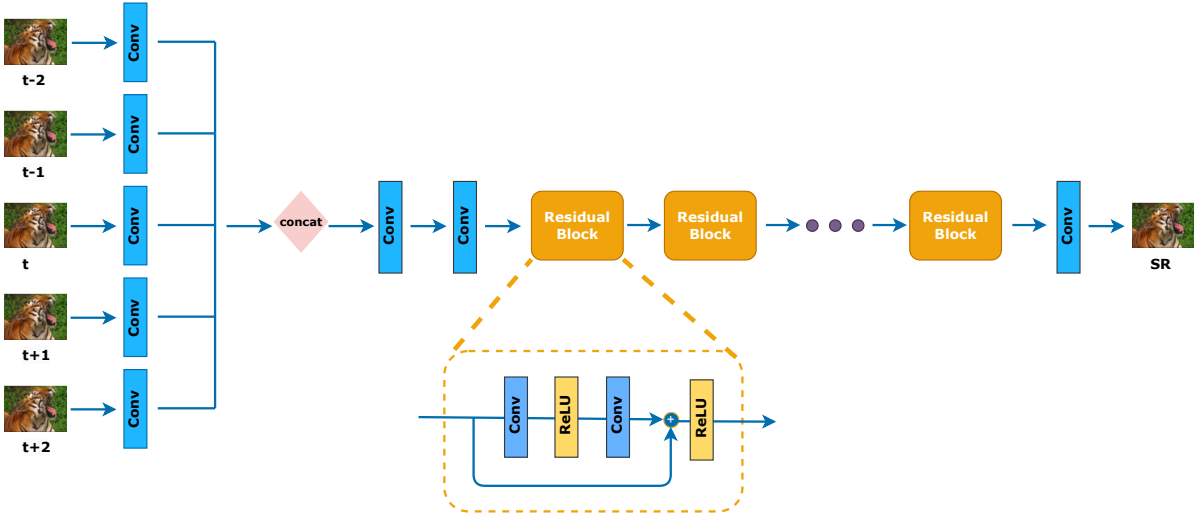

**Figure 6:** The VSRResFeatGAN's generator architecture. In the beginning, 5 conv operations are applied separately on each input bicubic interpolated frame at time t-2, t-1, t, t+1, t+2. The resulting feature maps are then concatenated together, and the combined feature map will be inputted into the following two conv layers and 15 residual blocks. Inside each residual block, there are two conv layers, each followed by a ReLU layer, and the input feature map is added to the output feature map to obtain the final output of the residual block. In the end, one more conv operation is conducted to obtain the final SR frame output at time t.[16]

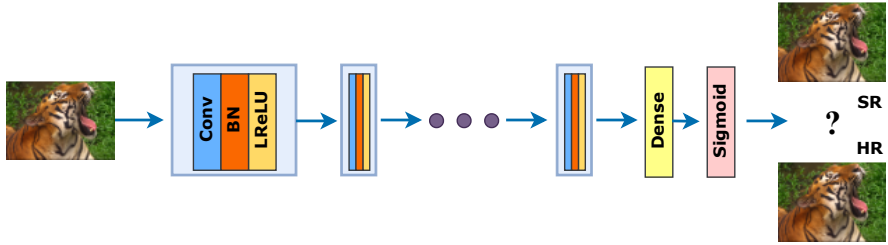

**Figure 7:** The VSRResFeatGAN's discriminator architecture. It consists of conv, batch normalization, and Leaky ReLU layers along the way, followed by one fully connected layer and a sigmoid activation function to obtain a probability. [16]

The original training loss used to train the ESRGAN contains the adversarial(GAN) loss, the perceptual loss, and the pixel loss [15]:

$$Loss^{esr} = \alpha L_{GAN} + L_{percep} + L_{pixel}, \qquad (16)$$

and the pixel loss used here is:

$$l_w^1(x, G_\theta(y), \beta\mathbf{1}). \qquad (17)$$

It follows the definition from Equation 8, where $x$ is the ground truth HR image, $y$ is the corresponding LR image, and it is also the input of generator network $G$ with trainable parameter $\theta$. Therefore $G_\theta(y)$ represents the output SR image. This pixel loss evaluates the L1 norm distance between the super-resolved image $G_\theta(y)$ and the ground truth image $x$ in the pixel space, and all elements in the images are equally weighted. The coefficients assigned to the GAN and pixel loss terms are represented as $\alpha$ and $\beta$, respectively. The coefficient for the perceptual loss term is fixed to 1.

The new training loss we used to retrain the ESRGAN replaces the original pixel loss with the proposed SA one and remains the same adversarial loss and perceptual loss

used in [15]. Therefore, the final SA training loss we used to train the model is defined as:

$$Loss^{SA-esr} = \alpha L_{GAN} + L_{percep} + L_{SA-pixel}, \qquad (18)$$

and the SA pixel loss used here is:

$$l_w^1(x, G_\theta(y), \beta_1\mathbf{1} + \beta_2 W(x)), \qquad (19)$$

which follows the definition of SA pixel loss from Equation 12. We name the model trained with this SA training loss (Equation 18) the SA-ESRGAN.

2) SA loss for video super-resolution model: VSRResFeat-GAN model also contains a generator and a discriminator. The generator's architecture is shown in Figure 6. Since the VSRResFeatGAN is used for video super-resolution task, it also takes temporal information in. Rather than using a single frame, its generator takes five consecutive frames as input, which are the bicubic interpolation on the LR frames at times t-2, t-1, t, t+1, and t+2. The resulting output is the corresponding SR frame at the central time instance, t. The discriminator's architecture is shown in Figure 7. Its input is either an HR frame or an SR frame generated from the generator, and its output is the probability of whether the input is a real HR frame or not. More details can be found in [16].

The original training loss used to train the VSRResFeat-GAN also contains three terms: the adversarial(GAN) loss, the perceptual loss, and the pixel loss [16]:

$$Loss^{vsr} = \alpha_1 L_{GAN} + \alpha_2 L_{percep} + L_{pixel}, \quad (20)$$

and the pixel loss used here is:

$$\gamma_w(x, G_\theta(Y), \beta \mathbf{1}), \quad (21)$$

it follows the definition from Equation 9, where $x$ is the ground truth HR frame. $Y$ is the bicubic-interpolated frames sequence, and it is also the input of the generator network $G$ with trainable parameter $\theta$, and $G_\theta(Y)$ represents the output SR frame at time t. This pixel loss evaluates the Chabonnier norm distance between the super-resolved frame $G_\theta(Y)$ and the ground truth frame $x$ in the pixel space, and all elements in the images are equally weighted. $\alpha_1$, $\alpha_2$, and $\beta$ are the coefficients for GAN, perceptual, and pixel loss terms.

Like what we have done with the SISR model (ESRGAN), the new training loss we used to retrain the VSRResFeatGAN replaces the original pixel loss with the corresponding SA one and remains the same adversarial loss and perceptual loss used in [16]. Therefore, the final SA training loss we used to train the model is defined as:

$$Loss^{SA-vsr} = \alpha_1 L_{GAN} + \alpha_2 L_{percep} + L_{SA-pixel}, \quad (22)$$

and the SA pixel loss used here is:

$$\gamma_w(x, G_\theta(Y), \beta_1 \mathbf{1} + \beta_2 W(x)), \quad (23)$$

which aligns with the definition of the SA pixel loss from Equation 13. We name the model trained with this SA training loss (Equation 22) the SA-VSRResFeatGAN.

## IV. EXPERIMENTS

This section shows the details of training the Single-image SR model (ESRGAN and SA-ESRGAN) and the video super-resolution model (VSRResFeatGAN and SA-VSRResFeatGAN).

### A. Datasets

The training dataset used for ESRGAN [15] and SA-ESRGAN is DIV2K [39], and the validation dataset used for them is Set14 [40]. The dataset used for VSRResFeatGAN [16] and SA-VSRResFeatGAN is Myanmar video dataset [41]. Myanmar video dataset contains 59 video sequences, and we take 53 of them to make the training (80%) and validation (20%) datasets. The rest 6 are used for testing. We followed the same data generation approaches described in [15] and [16], respectively. All experiments are performed under the scale factor 4 between LR and HR image pairs.

|  | PSNR | SSIM | LPIPS |
|---|---|---|---|
| ESRGAN | 29.13 | 0.8445 | 0.0398 |
| SA-ESRGAN (lv) | 29.23 | 0.8444 | 0.0387 |
| SA-ESRGAN (canny) | **29.75** | **0.8561** | **0.0357** |

TABLE I: Metrics comparison between ESRGAN, SA-ESRGAN (local-variance), and SA-ESRGAN (canny). The results are evaluated on Myanmar testing Dataset for scale factor 4.

|  | PSNR | SSIM | LPIPS |
|---|---|---|---|
| ESRGAN | 21.53 | 0.6349 | 0.1022 |
| SA-ESRGAN (lv) | 22.14 | 0.6581 | **0.0953** |
| SA-ESRGAN (canny) | **22.32** | **0.6674** | 0.0955 |

TABLE II: Metrics comparison of ESRGAN, SA-ESRGAN (local-variance), and SA-ESRGAN (canny) in PSNR, SSIM, and LPIPS metrics. For the LPIPS, smaller is better. The results are evaluated on VidSet4 Dataset for scale factor 4.

### B. Training Process

To train the ESRGAN and SA-ESRGAN models, we first initialize their generators with the PSNR-oriented pre-trained model, provided by the author of ESRGAN [15]. Such initialization facilitates proper convergence during the following GAN-based training phase. Within the GAN-based training, the ESRGAN model's generator is optimized using the loss function defined in Equation 16, while the SA-ESRGAN model's generator is optimized using the loss function defined in Equation 18. For both models, the learning rates for the generator and discriminator are initialized to $10^{-4}$, then halved at [50k, 100k, 200k, 300k] iterations ($k = 10^3$). To guarantee fair comparisons, we set the maximum training iteration number as 500k and use the minimum validation loss as the termination criterion for both models. The validation loss is computed by Equation 16 and Equation 18 on the validation dataset, respectively. The optimizer we use is Adam [42] with batch size 16. The generator and discriminator are updated alternately.

The training processes of the VSRResFeatGAN and SA-VSRResFeatGAN models are similar. We first initialize the generator with the PSNR-oriented pre-trained model, provided by the author of VSRResFeatGAN [16]. In the following GAN-based training process, for the VSRResFeatGAN model, the generator is trained using the loss function in Equation 20, and for the SA-VSRResFeatGAN model, the generator is trained using the loss function in Equation 22. For both models, the learning rates for the generator and discriminator are set to $10^{-4}$. We set the maximum training epoch number to 40 and also use the minimum validation loss as the training termination criterion for both models. The validation loss is computed using Equation 20 and Equation 22 on the validation dataset, respectively. The optimizer we use is Adam with batch size 64. The generator and discriminator are updated alternately.

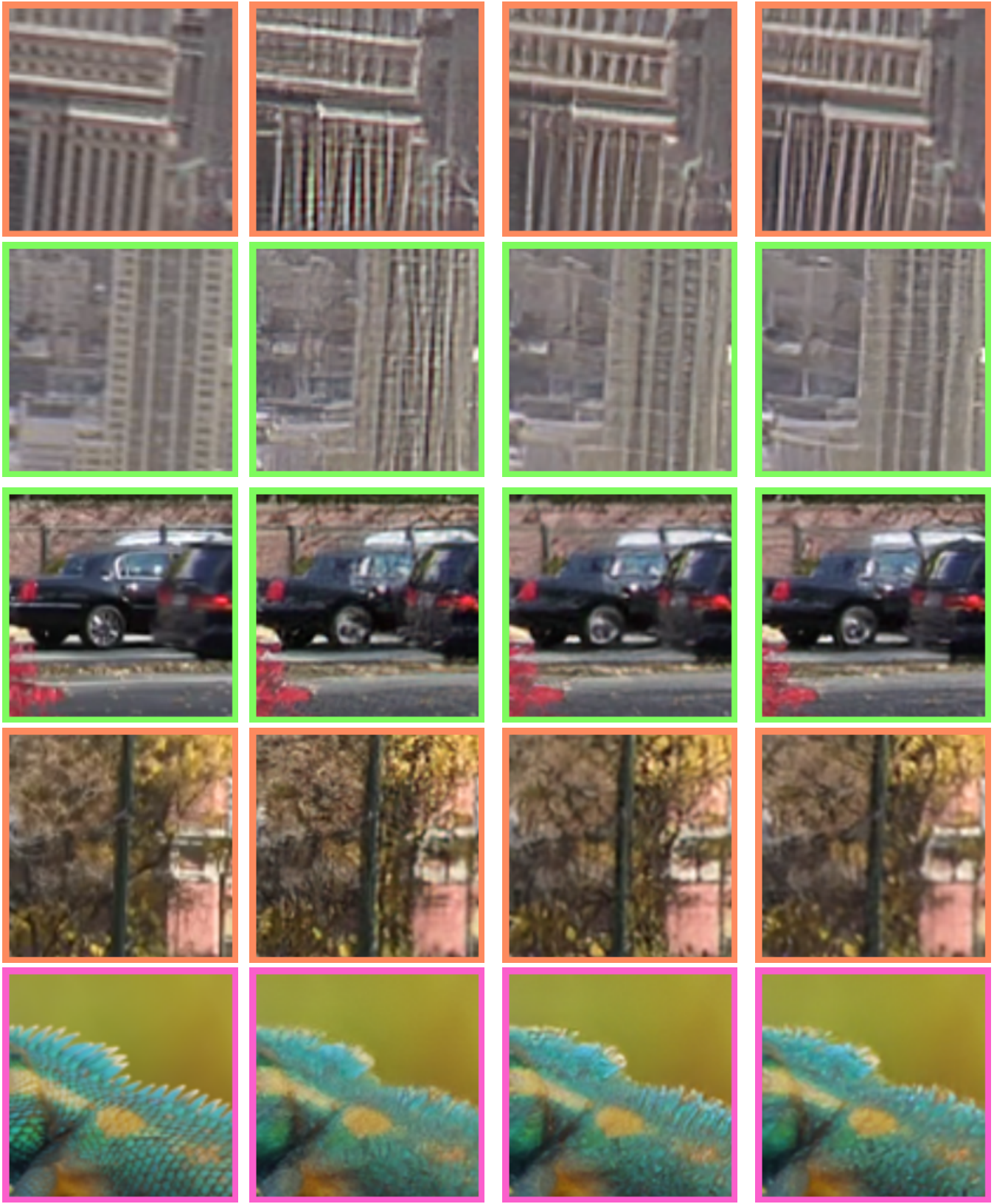

|Ground truth|ESRGAN|SA-ESRGAN (lv)|SA-ESRGAN (canny)|

**Figure 8:** Qualitative comparison of ESRGAN, SA-ESRGAN(lv), and SA-ESRGAN(canny).

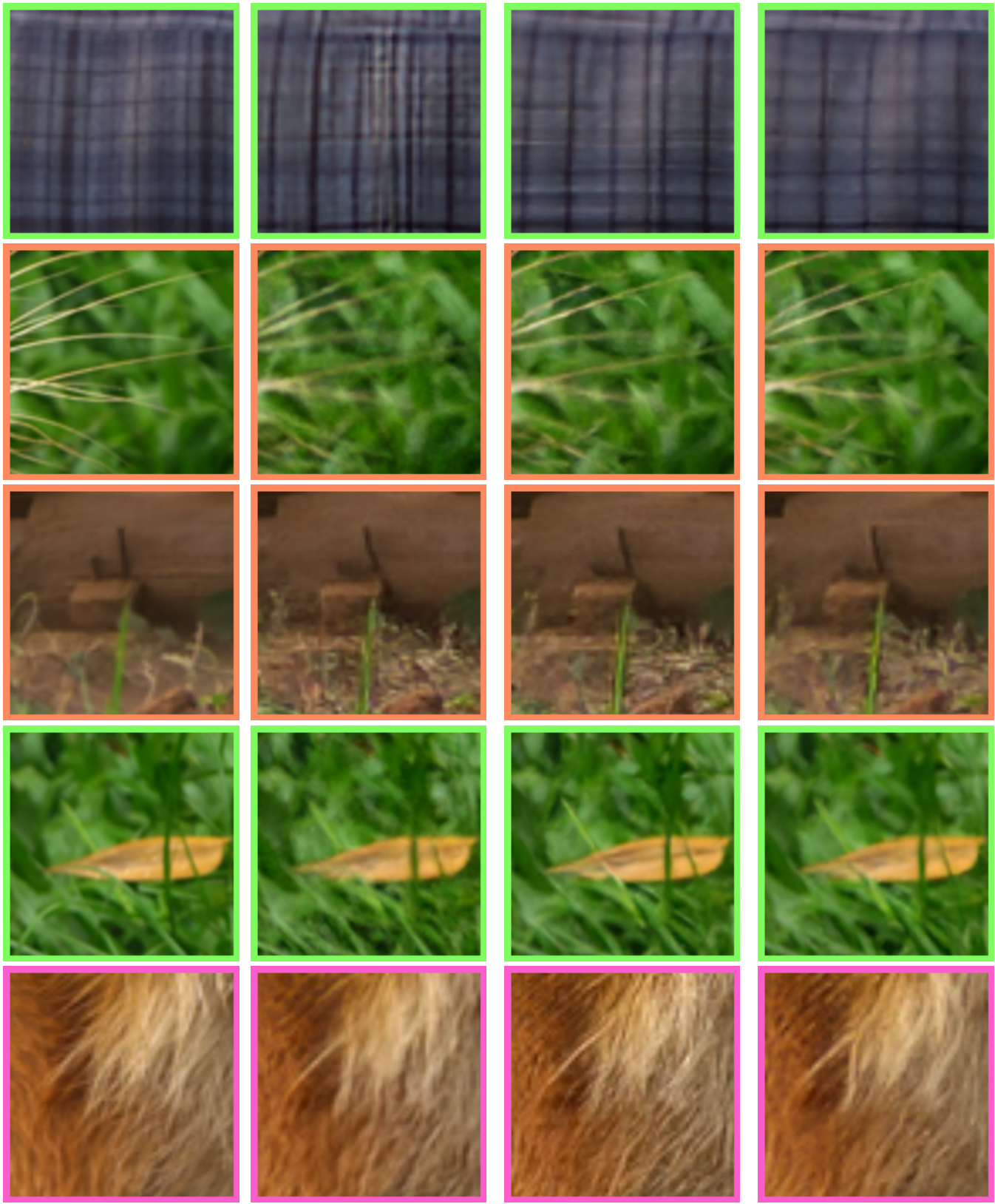

**Figure 9:** Qualitative comparison of ESRGAN, SA-ESRGAN(lv), and SA-ESRGAN(canny).

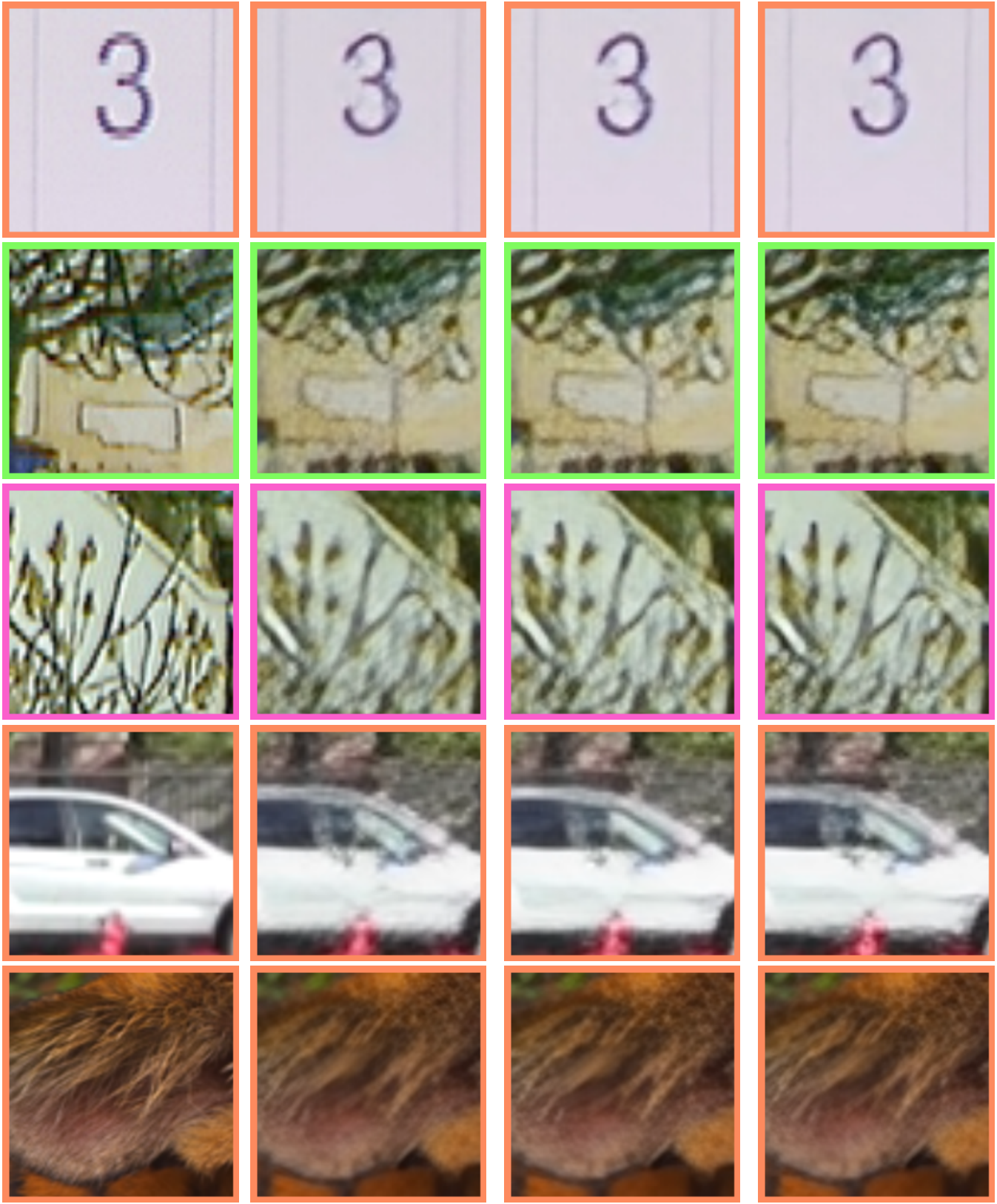

Ground truth       VSRResFeatGAN       SA-VSRResFeatGAN (lv)       SA-VSRResFeatGAN (canny)

**Figure 10:** Qualitative comparison of VSRResFeatGAN, SA-VSRResfeatGAN(lv), and SA-VSRResfeatGAN(canny).

## C. Hyper-parameters for the Loss Functions

For training the ESRGAN and VRResFeatGAN models, we choose the same coefficients used in [15] and [16] respectively: $\alpha = 0.005$, $\beta = 0.01$ in Equation 16 and 17 when training the ESRGAN model. $\alpha_1 = 0.001$, $\alpha_2 = 0.998$ and $\beta = 0.001$ in Equation 20 and 21 when training the VSRResFeatGAN model.

During the training of the SA-ESRGAN and SA-VSRResFeatGAN models, we have determined that when the SA loss component constitutes approximately 15% of the total loss value, it exhibits the ability to exert a substantial and appropriate influence on the generator, leading to the production of sharper and more distinct edges in the generated outcomes. Based on this observation, we have set the coefficients for the loss terms as follows: for SA-ESRGAN model, the generator is trained employing the loss outlined in Equation 18 and Equation 19, with $\alpha_1 = 0.005$, $\beta_1 = 0.01$, and $\beta_2 = 20$. For SA-VSRResFeatGAN model, its generator is trained using the loss function outlined in Equation 22 and Equation 23, with $\alpha_1 = 0.001$, $\alpha_2 = 0.998$ and $\beta_1 = 0.001$, and we set $\beta_2 = 5$ when using the local variance edge extraction method to compute weight matrix $W$, while $\beta_2 = 1.5$ when using the canny edge extraction method. In both models, $\beta_2$ governs the influence of the SA loss component; its values are decided experimentally so that the SA loss part took up around 15% of the total training loss, respectively.

## V. Results

In this section, we show the impact of the proposed SA loss on the outcomes. By doing so, we will provide both qualitative and quantitative comparisons between the SISR model and the VSR model, trained both with and without the SA loss, i.e., ESRGAN vs. SA-ESRGAN and VRResFeatGAN vs. SA-VSRResFeatGAN.

In Table II and I, we show the comparison of ESRGAN and SA-ESRGAN in terms of PSNR, SSIM, and the Learned Perceptual Image Patch Similarity (LPIPS) [43] metrics for VidSet4 and Myanmar test datasets. LPIPS is a new standard to compare the perceptual similarity between a reference image and a distorted one with a CNN. The author found out this perceptual distance provides results consistent with human judgment [44]. We also observed that it was consistent with the human's opinion regarding the sharpness of the produced super-resolved images. For the SA-ESRGAN model, based on the edge detection algorithm used to compute the weight matrix $W(x)$ in Equation 19, we categorize the model into SA-ESRGAN (lv) and SA-ESRGAN (canny), i.e., they are trained with the local variance-based SA loss and the Canny detection–based SA loss, respectively. We can see that for the Vidset 4 test dataset (Table II), the SA-ESRGAN models surpass the ESRGAN model across all metrics. Similar trends emerge for the Myanmar test dataset (Table I), where all SA-ESRGAN models outperform the ESRGAN model in terms of the LPIPS metric. This fact verifies our proposal that using the SA loss for the training can help guide the SISR model to generate images with higher perceptual quality. The qualitative comparisons are shown in Figure 8-9. We can see that the SA-ESRGAN yields more precise super-resolution of edges and fine details with reduced noise than the ESRGAN.

Similarly, for the VSR models, we show the quantitative comparison of VSRResFeatGAN and SA-VSRResFeatGAN in Table III and IV. Again, SA-VSRResFeatGAN models surpass the VSRResFeatGAN model almost for all the metrics' values. Especially for the LPIPS metric, it consistently improves after SA loss is used, which verifies our proposal that the use of the SA loss during training can guide the VSR model towards generating images with higher perceptual quality and better visual feelings. The qualitative comparisons are shown in Figure 10. We can observe that the SA-VSRResFeatGAN generates more accurate super-resolved edges and fine details with less noise compared with the VSRResFeatGAN.

|  | PSNR | SSIM | LPIPS |
|---|---|---|---|
| **VSRResFeatGAN** | 24.71 | 0.7199 | 0.1045 |
| **SA-VSRResFeatGAN (lv)** | **25.09** | **0.7344** | **0.0992** |
| **SA-VSRResFeatGAN (canny)** | 24.74 | 0.7192 | 0.1027 |

TABLE III: Metrics comparison of VSRResFeatGAN, SA-VSRResFeatGAN (local-variance), and SA-VSRResFeatGAN (canny) in terms of PSNR, SSIM, and LPIPS metrics. For the LPIPS, smaller is better. The results are evaluated on VidSet4 Dataset for scale factor 4.

|  | PSNR | SSIM | LPIPS |
|---|---|---|---|
| **VSRResFeatGAN** | 32.22 | 0.8887 | 0.0545 |
| **SA-VSRResFeatGAN (lv)** | **32.48** | **0.8918** | 0.0540 |
| **SA-VSRResFeatGAN (canny)** | 32.36 | 0.8890 | **0.0539** |

TABLE IV: Metrics comparison between VSRResFeatGAN, SA-VSRResFeatGAN (local-variance), and SA-VSRResFeatGAN (canny). The results are evaluated on Myanmar testing Dataset for scale factor 4.

## VI. Conclusion

In this paper, we proposed the spatially adaptive(SA) loss for training SR models. It integrates spatial information extracted from input images/video frames into the training loss. Specifically, we focus on edges as a pivotal spatial component. Models trained using the SA loss demonstrate the capacity to generate super-resolved images/video frames with enhanced perceptual quality.

We have shown that the proposed method is independent of the edge information extraction approach, as long as it effectively detects the edges. By employing extracted edge information to formulate the SA loss and subsequently retraining the SR model using this loss, we could consistently observe improved results with better metrics values and sharper edges, improved reconstruction of fine details, and a significant decrease in noise. Furthermore, we have presented that the SA loss could help both SISR and VSR GAN models consistently achieve better performance. This substantiates the adaptability and efficacy of our proposed SA loss across broader scenarios, steering SR models toward the production of visually pleasing outcomes.

# REFERENCES

[1] A. Lucas, M. Iliadis, R. Molina, and A. K. Katsaggelos, "Using deep neural networks for inverse problems in imaging: beyond analytical methods," *IEEE Signal Processing Magazine*, vol. 35, no. 1, pp. 20–36, 2018.

[2] Z. Feng, W. Zhang, S. Liang, and Q. Yu, "Deep video super-resolution using hybrid imaging system," *IEEE Transactions on Circuits and Systems for Video Technology*, 2023.

[3] H. Wu, J. Gui, J. Zhang, J. T. Kwok, and Z. Wei, "Feedback pyramid attention networks for single image super-resolution," *IEEE Transactions on Circuits and Systems for Video Technology*, 2023.

[4] F. Zhang, G. Chen, H. Wang, J. Li, and C. Zhang, "Multi-scale video super-resolution transformer with polynomial approximation," *IEEE Transactions on Circuits and Systems for Video Technology*, 2023.

[5] R. Chen and Y. Zhang, "Learning dynamic generative attention for single image super-resolution," *IEEE Transactions on Circuits and Systems for Video Technology*, vol. 32, no. 12, pp. 8368–8382, 2022.

[6] M. C. Zerva and L. P. Kondi, "Video super-resolution using plug-and-play priors," *IEEE Access*, 2024.

[7] C. Dong, C. C. Loy, K. He, and X. Tang, "Learning a deep convolutional network for image super-resolution," in *European conference on computer vision*. Springer, 2014, pp. 184–199.

[8] ——, "Image super-resolution using deep convolutional networks," *IEEE transactions on pattern analysis and machine intelligence*, vol. 38, no. 2, pp. 295–307, 2015.

[9] A. Kappeler, S. Yoo, Q. Dai, and A. K. Katsaggelos, "Video super-resolution with convolutional neural networks," *IEEE Transactions on Computational Imaging*, vol. 2, no. 2, pp. 109–122, 2016.

[10] J. Kim, J. K. Lee, and K. M. Lee, "Accurate image super-resolution using very deep convolutional networks," in *Proceedings of the IEEE conference on computer vision and pattern recognition*, 2016, pp. 1646–1654.

[11] B. Lim, S. Son, H. Kim, S. Nah, and K. Mu Lee, "Enhanced deep residual networks for single image super-resolution," in *Proceedings of the IEEE conference on computer vision and pattern recognition workshops*, 2017, pp. 136–144.

[12] I. J. Goodfellow, J. Pouget-Abadie, M. Mirza, B. Xu, D. Warde-Farley, S. Ozair, A. Courville, and Y. Bengio, "Generative adversarial networks," *arXiv preprint arXiv:1406.2661*, 2014.

[13] M. S. Sajjadi, B. Scholkopf, and M. Hirsch, "Enhancenet: Single image super-resolution through automated texture synthesis," in *Proceedings of the IEEE International Conference on Computer Vision*, 2017, pp. 4491–4500.

[14] C. Ledig, L. Theis, F. Huszár, J. Caballero, A. Cunningham, A. Acosta, A. Aitken, A. Tejani, J. Totz, Z. Wang *et al.*, "Photo-realistic single image super-resolution using a generative adversarial network," in *Proceedings of the IEEE conference on computer vision and pattern recognition*, 2017, pp. 4681–4690.

[15] X. Wang, K. Yu, S. Wu, J. Gu, Y. Liu, C. Dong, Y. Qiao, and C. Change Loy, "Esrgan: Enhanced super-resolution generative adversarial networks," in *Proceedings of the European Conference on Computer Vision (ECCV) Workshops*, 2018, pp. 0–0.

[16] A. Lucas, S. Lopez-Tapia, R. Molina, and A. K. Katsaggelos, "Generative adversarial networks and perceptual losses for video super-resolution," *IEEE Transactions on Image Processing*, vol. 28, no. 7, pp. 3312–3327, 2019.

[17] E. Pérez-Pellitero, M. S. Sajjadi, M. Hirsch, and B. Schölkopf, "Photorealistic video super resolution," in *Workshop and Challenge on Perceptual Image Restoration and Manipulation (PIRM) at the 15th European Conference on Computer Vision (ECCV)*, 2018.

[18] M. Zhang and Q. Ling, "Supervised pixel-wise gan for face super-resolution," *IEEE Transactions on Multimedia*, vol. 23, pp. 1938–1950, 2020.

[19] S. López-Tapia, A. Lucas, R. Molina, and A. K. Katsaggelos, "A single video super-resolution gan for multiple downsampling operators based on pseudo-inverse image formation models," *Digital Signal Processing*, vol. 104, p. 102801, 2020.

[20] A. Lugmayr, M. Danelljan, F. Yu, L. Van Gool, and R. Timofte, "Normalizing flow as a flexible fidelity objective for photo-realistic super-resolution," in *Proceedings of the IEEE/CVF Winter Conference on Applications of Computer Vision*, 2022, pp. 1756–1765.

[21] J. Guerreiro, P. Tomás, N. Garcia, and H. Aidos, "Super-resolution of magnetic resonance images using generative adversarial networks," *Computerized Medical Imaging and Graphics*, p. 102280, 2023.

[22] J. Song, H. Yi, W. Xu, X. Li, B. Li, and Y. Liu, "Esrgan-dp: Enhanced super-resolution generative adversarial network with adaptive dual perceptual loss," *Heliyon*, vol. 9, no. 4, 2023.

[23] K. Jiang, Z. Wang, P. Yi, G. Wang, T. Lu, and J. Jiang, "Edge-enhanced gan for remote sensing image superresolution," *IEEE Transactions on Geoscience and Remote Sensing*, vol. 57, no. 8, pp. 5799–5812, 2019.

[24] X. Wang, K. Yu, C. Dong, and C. C. Loy, "Recovering realistic texture in image super-resolution by deep spatial feature transform," in *Proceedings of the IEEE conference on computer vision and pattern recognition*, 2018, pp. 606–615.

[25] X. Wu, A. Lucas, S. Lopez-Tapia, X. Wang, Y. H. Kim, R. Molina, and A. K. Katsaggelos, "Semantic prior based generative adversarial network for video super-resolution," in *2019 27th European Signal Processing Conference (EUSIPCO)*. IEEE, 2019, pp. 1–5.

[26] J. Zhao, Y. Ma, F. Chen, E. Shang, W. Yao, S. Zhang, and J. Yang, "Sa-gan: A second order attention generator adversarial network with region aware strategy for real satellite images super resolution reconstruction," *Remote Sensing*, vol. 15, no. 5, p. 1391, 2023.

[27] Y. Zhang, K. Li, K. Li, L. Wang, B. Zhong, and Y. Fu, "Image super-resolution using very deep residual channel attention networks," in *Proceedings of the European conference on computer vision (ECCV)*, 2018, pp. 286–301.

[28] N. Ahn, B. Kang, and K.-A. Sohn, "Fast, accurate, and lightweight super-resolution with cascading residual network," in *Proceedings of the European Conference on Computer Vision (ECCV)*, 2018, pp. 252–268.

[29] X. Wang, K. C. Chan, K. Yu, C. Dong, and C. Change Loy, "Edvr: Video restoration with enhanced deformable convolutional networks," in *Proceedings of the IEEE/CVF Conference on Computer Vision and Pattern Recognition Workshops*, 2019, pp. 0–0.

[30] J. Johnson, A. Alahi, and L. Fei-Fei, "Perceptual losses for real-time style transfer and super-resolution," in *European conference on computer vision*. Springer, 2016, pp. 694–711.

[31] A. Bulat and G. Tzimiropoulos, "Super-fan: Integrated facial landmark localization and super-resolution of real-world low resolution faces in arbitrary poses with gans," in *Proceedings of the IEEE Conference on Computer Vision and Pattern Recognition*, 2018, pp. 109–117.

[32] S. Vasu, N. Thekke Madam, and A. Rajagopalan, "Analyzing perception-distortion tradeoff using enhanced perceptual super-resolution network," in *Proceedings of the European Conference on Computer Vision (ECCV) Workshops*, 2018, pp. 0–0.

[33] J.-H. Choi, J.-H. Kim, M. Cheon, and J.-S. Lee, "Deep learning-based image super-resolution considering quantitative and perceptual quality," *Neurocomputing*, vol. 398, pp. 347–359, 2020.

[34] M. Cheon, J.-H. Kim, J.-H. Choi, and J.-S. Lee, "Generative adversarial network-based image super-resolution using perceptual content losses," in *Proceedings of the European Conference on Computer Vision (ECCV) Workshops*, 2018, pp. 0–0.

[35] S. N. Efstratiadis and A. K. Katsaggelos, "Adaptive iterative image restoration with reduced computational load," *Optical engineering*, vol. 29, no. 12, pp. 1458–1468, 1990.

[36] J. Canny, "A computational approach to edge detection," *IEEE Transactions on pattern analysis and machine intelligence*, no. 6, pp. 679–698, 1986.

[37] G. Bradski, "The OpenCV Library," *Dr. Dobb's Journal of Software Tools*, 2000.

[38] K. Simonyan and A. Zisserman, "Very deep convolutional networks for large-scale image recognition," *arXiv preprint arXiv:1409.1556*, 2014.

[39] E. Agustsson and R. Timofte, "Ntire 2017 challenge on single image super-resolution: Dataset and study," in *Proceedings of the IEEE Conference on Computer Vision and Pattern Recognition Workshops*, 2017, pp. 126–135.

[40] R. Zeyde, M. Elad, and M. Protter, "On single image scale-up using sparse-representations," in *International conference on curves and surfaces*. Springer, 2010, pp. 711–730.

[41] "Myanmar 60p, harmonic inc. (2014)," http://www.harmonicinc.com/resources/videos/4k-video-clip-center.

[42] D. P. Kingma and J. Ba, "Adam: A method for stochastic optimization," *arXiv preprint arXiv:1412.6980*, 2014.

[43] R. Zhang, P. Isola, A. A. Efros, E. Shechtman, and O. Wang, "The unreasonable effectiveness of deep features as a perceptual metric," in *Proceedings of the IEEE conference on computer vision and pattern recognition*, 2018, pp. 586–595.

[44] K. Zhang, W. Zuo, S. Gu, and L. Zhang, "Learning deep cnn denoiser prior for image restoration," in *Proceedings of the IEEE conference on computer vision and pattern recognition*, 2017, pp. 3929–3938.